\documentstyle[12pt]{article}

\begin{document}
\begin{center}
{\Large \bf The universal potential fluctuations}
\bigskip

{\large D.L.~Khokhlov}
\smallskip

{\it Sumy State University, R.-Korsakov St. 2, \\
Sumy 40007, Ukraine\\
E-mail: khokhlov@cafe.sumy.ua}
\end{center}

\begin{abstract}
It is assumed the existence of the universal potential fluctuations
valid for all scales in the universe
which follow the fractal law $\delta_U=(\Delta r/r)^2$.
The value of the universal potential fluctuations is determined
from the data on anisotropy of the cosmic microwave
background on degree scales. It is suggested
to test the existence of the universal potential fluctuations
in the laboratory experiment.
\end{abstract}

There are two independent arguments that
the universe can be viewed as a fractal
with the law
\begin{equation}
m(<r) \propto r^2
\label{eq:mrd}
\end{equation}
where $m(<r)$ is the mass restricted within radius $r$.
With the use of heuristic conjecture Dirac~\cite{Dir} obtained
the fractal relation between the microworld
scale and the universe scale
\begin{equation}
\frac{m_U}{m_{mic}}=\left(\frac{r_U}{r_{mic}}\right)^{2}
\label{eq:Rr}
\end{equation}
where $m_U$ and $r_U$ are the mass and the radius of the universe,
$m_{mic}$ and $r_{mic}$ are the mass and the radius of microparticle.
Pietronero et al.~\cite{Pi}
claimed that galaxies have a fractal distribution
\begin{equation}
N(<r) \propto r^2
\label{eq:monop}
\end{equation}
on all scales up to the deepest scales probed until now
$1000\ h^{-1}\ {\rm Mpc}$.
Thus it is natural
to assume that the fractal law $m(<r) \propto r^2$ extends
on all scales from the microworld scale to the universe scale.

Let us assume that there exist universal potential fluctuations
valid for all scales in the universe
\begin{equation}
\delta_{U}=\frac{\Delta m}{m}=
\left(\frac{\Delta r}{r}\right)^2.
\label{eq:dl}
\end{equation}
That is the universal potential fluctuations follow the above
fractal law.
The universal potential fluctuations can be registered as
angular fluctuations of the electromagnetic radiation.
Let us assume that the universal potential fluctuations
cause anisotropy of the cosmic microwave
background (CMB) on degree scales.
The peak of the temperature fluctuations in
the CMB spectrum $\Delta T/T=3\times 10^{-5}$ is seen at the
multipole $\ell_{eff}=260$~\cite{Han}. These values can be
bounded by the relation
\begin{equation}
\frac{\Delta T}{T}\sim 2\ell_{eff}^{-2}.
\label{eq:Dt}
\end{equation}
Let us assume that the multipole $\ell_{eff}$ relates with
the universal potential fluctuations as
\begin{equation}
\delta_{U}=\ell_{eff}^{-2}.
\label{eq:du}
\end{equation}
The value $\ell_{eff}=260$ yields the value
$\delta_U=1.48 \times 10^{-5}$.

It is interesting to test the existence of
the universal potential fluctuations in the laboratory experiment.
The scheme of the experiment is the following. Two identical
sources of the electromagnetic radiation are placed at the
distance $\Delta r$ between them and at the distance $r$ from the
detector. The universal potential fluctuations should cause
fluctuations of the temperature of the electromagnetic radiation
$\Delta T/T\sim 3\times 10^{-5}$ at the angle
$\Delta r/r \sim 1/260$.

\end{document}